\documentclass[11pt]{article}
\usepackage[final]{graphics}
\usepackage{amssymb,amsfonts}

\font\tenmsa=msam10
\font\sevenmsa=msam7
\font\fivemsa=msam5
\font\tenmsb=msbm10
\font\sevenmsb=msbm7
\font\fivemsb=msbm5
\newfam\msafam
\newfam\msbfam
\textfont\msafam=\tenmsa  \scriptfont\msafam=\sevenmsa
\scriptscriptfont\msafam=\fivemsa
\textfont\msbfam=\tenmsb  \scriptfont\msbfam=\sevenmsb
\scriptscriptfont\msbfam=\fivemsb

\global\mathchardef\lesssim "142E

\newcommand{\slL}{\raise.15ex\hbox{$/$}\kern-.53em\hbox{$L$}}
\newcommand{\slP}{\raise.15ex\hbox{$/$}\kern-.53em\hbox{$P$}}
\newcommand{\slR}{\raise.15ex\hbox{$/$}\kern-.53em\hbox{$R$}}
\newcommand{\slQ}{\raise.15ex\hbox{$/$}\kern-.53em\hbox{$Q$}}
\newcommand{\slK}{\raise.15ex\hbox{$/$}\kern-.53em\hbox{$K$}}
\newcommand{\slSigma}{\raise.15ex\hbox{$/$}\kern-.53em\hbox{$\Sigma$}}
\newcommand{\slcalP}{\raise.15ex\hbox{$/$}\kern-.63em\hbox{$\cal P$}}

\newcommand{\be}{\begin{equation}}
\newcommand{\ee}{\end{equation}}     
\newcommand{\bea}{\begin{eqnarray}}
\newcommand{\ena}{\end{eqnarray}}

\def\build#1\over#2{\mathrel{\mathop{\kern 0pt#1}\limits_{#2}}}
\def\oldstyle#1{$\mit #1$}

\font\tenimbf=cmmib10 at 12pt
\font\sevenimbf=cmmib10 at 7pt
\font\fiveimbf=cmmib10 at 5pt
\newfam\imbf
\textfont\imbf=\tenimbf
\scriptfont\imbf=\sevenimbf
\scriptscriptfont\imbf=\fiveimbf

\begin{document}
\begin{titlepage}
\title{\begin{center}
{\Huge LAPTH}
\end{center}
\vspace{5 mm}
\hrule
\vspace{20mm}
\bf{Finite temperature world-line formalism and 
  analytic continuation}}
\author{
H.~Zaraket}
\maketitle

\begin{center}
Lebanese University Faculty of Sciences (I)\\
Hadeth-Beirut, Lebanon\\
and \\
Laboratoire de Physique Th\'eorique LAPTH,\\
UMR 5108 du CNRS, associ\'ee \`a l'Universit\'e de Savoie,\\
BP110, F-74941, Annecy le Vieux Cedex, France
\end{center}

\begin{abstract} 
In vacuum, the world-line formalism is an efficient tool for calculating observables in 
the presence of arbitrary constant external fields. The natural frame of this 
formalism is the Euclidean space. At finite temperature the analytic 
continuation to Minkowski 
space is a subtle task. We study the two point function in scalar QED, and we 
figure out the problem of analytic continuation giving a possible solution 
for it. We also show, in contrast to what was claimed in the 
literature, that a translationally invariant world-line Green's function could 
be used at finite temperature.    
\end{abstract}
   \vskip 4mm
\centerline{\hfill LAPTH--866/01}
\vfill
\thispagestyle{empty}
\end{titlepage}

\section{Introduction}
The study of the effect of an external field on a thermalised system is of 
great importance. It is suggested \cite{Gusyn1} that a strong magnetic field 
could have existed during the early stages of the Universe. This would have 
affected the 
features of the electroweak phase transition. The effect of a magnetic field 
on a relativistic QED or QCD 
plasma has been investigated in the literature \cite{AlexaFK1}. As this magnetic field has been generated by the 
phase transition, it is not expected to have  
overwhelmed thermal effect \cite{AlexaFK1}. On the other hand an 
electric field is expected to have primordial importance due to its 
strong interaction with matter and the long period over which it last. The 
effect of an electric field at finite temperature has not yet been achieved. \\
For our study we assume that we have thermal and chemical equilibrium. We 
take the chemical potential to be zero.\\ 
The world-line formalism is a well adapted tool to incorporate an external 
magnetic field when calculating correlation functions. In vacuum, this 
formalism allows to obtain the n-point Green's functions in the presence of 
an external field 
from those in the absence of such field. In this paper we show that such 
an advantage does not persist at finite temperature due to the problem of 
analytic continuation. The starting point of the world-line 
formalism is the Euclidean space, where the convergence of integrals is well 
controlled. The simple derivation of the world-line effective action in 
Euclidean space is balanced by 
the subtle analytic continuation to Minkowski space at finite temperature.
 This makes 'Wick's rotation' very tricky. The example of the two point 
Green's function in scalar QED traces the problem of 
analytic continuation to the use of Feynman parametrization. The naive use of 
Feynman parametrization leads to wrong results. Hence the 
advantage of obtaining easily the two point function in the presence of an external 
field is lost. Therefore a direct calculation of the two point function, in 
the presence of an external field, should be performed.\\    
We recall that the subtleties discussed in the paper, are not applicable to 
thermodynamical functions (entropy, ...). Those functions have no 
external momentum and hence there is no analytic continuation to be done. 

The paper is organized as follows. In the first section we review the basic 
notations of the world-line formalism, and we show that a translationally 
invariant Green's function can be used at finite temperature. Then the two 
point function is calculated in the world-line formalism to illustrate the 
relation between this formalism and Feynman parameters. Finally the problem of 
analytic continuation is discussed and a possible solution is proposed.

\section{Notation and formalism}
\subsection{Zero temperature field theory}
The one-loop effective action in a first quantized path integral form 
is well known at zero temperature \cite{BernK1}. A finite temperature 
extension can be found in the literature \cite{MckeoR1}. Recently the Hard Thermal 
Loop effective action was derived in the world-line formalism \cite{VenugW1}. In this paper 
we use the same notation used in \cite{MckeoR1}. 
We do not intend here to give a complete derivation of the 
effective action but rather a sketch of the important steps leading to its 
final form. \\
At zero temperature we could proceed in the following way to get the 
effective action \cite{Stras1}:  
\begin{itemize}
\item In Euclidean space, the starting point is the standard 'Trlog(propagator operator)' 
formula \cite{Sterm1}:\\
{\it e.g.} for scalar field theory, of mass $m$ and potential $V(\phi)$, the one loop effective 
action is given by $$\Gamma_{\rm eff}\sim {\rm Tr}\log\left(-\square+m^2 + V^{''}(\phi)\right).$$    
\item Then we use the following formula 
$${\rm Tr}\log(A)= {\rm Tr} \int\limits^{\infty}_{0}\frac{dt}{t}\exp(-At)$$ 
to get a form similar to the path integral obtained for the evolution 
operator $\exp-(t-t^\prime)H$, where $H$ is the first quantized Hamiltonian, 
with a kinetic term $\sim \dot{x}^2/2$ and a potential term. 
\end{itemize}
By following these steps, the scalar QED one-loop effective action is 
found to be \cite{Schub1}
\begin{equation}
\Gamma_{\rm eff}[A]= \int\limits^{\infty}_{0}\frac{dt}{t}e^{-m^2 t}\int\limits_{x(t)=x(0)}{\cal D}x(\tau)\exp-\int\limits_{0}^{t}d\tau\left(\frac{\dot{x}^2}{4}+ie\dot{x}.A(x(\tau))\right) .
\label{eq:action}
\end{equation} 
The n-point Green's function can easily be derived by using the Fourier transform 
of the photon field $A^\mu(x)=\sum \epsilon_\mu^i e^{ik_i x}$. For the above 
effective action Wick's contraction is found to be  
$$<x|x^\mu(\tau_1)x^{\nu}(\tau_2)|x>=-g^{\mu\nu}G(\tau_1-\tau_2)$$ 
with the convention $g^{\mu\nu}=(+,+,+,+)$. The one dimension translationally 
invariant Green's function G is given by: 
\begin{equation}
G(\tau_1-\tau_2)=<x|-\frac{d^2}{d\tau^2}|x>= |\tau_1-\tau_2|-\frac{(\tau_1-\tau_2)^2}{t}.
\label{eq:green}
\end{equation} 
Deriving the effective action twice with respect to $\epsilon$ and using Wick's contraction, the two 
point function in D-dimension can easily be found:
\begin{eqnarray}
\Pi^{\mu\nu}(k)= -\frac{e^2}{2}\int\limits_0^\infty\frac{dt}{t}e^{-m^2 t}[4\pi t]^{-D/2}\int\limits_0^td\tau_1\int\limits_0^td\tau_2 [g^{\mu\nu}k^2-k^\mu k^\nu]\nonumber\\
\times\dot{G}^2(\tau_1-\tau_2)\exp(-G(\tau_1-\tau_2)k^2)\; .
\label{eq:zerotemp}
\end{eqnarray}
where $g^{\mu\nu}k^2-k^\mu k^\nu$ is the standard transverse tensorial structure of 
the two point Green's function.
  
\subsection{Finite temperature field theory two point function} 
We work in scalar QED, whoever in the world-line formalism 
it is straight forward to obtain spinor QED from scalar QED. In addition at high 
temperature we can neglect the mass of the scalar particle.\\  
At finite temperature one has to take into account the periodicity conditions.
 The boundary condition {$x(t)=x(0)$} in the path integral of 
Eq.~(\ref{eq:action}) changes to {$x_\mu(t)=x_\mu(0)+n\beta w_\mu$}, with 
$\beta=1/{\rm T}$ and $w_{\mu}=(1,0,0,0)$. We have to sum also over the 
winding number '$n$' \cite{MckeoR1}. Then the effective action at finite 
temperature is given by 
\begin{equation}
\Gamma_{\rm eff}^{\beta}[A]=\left.\sum\limits_{n=-\infty}^{\infty}\Gamma_{\rm eff}[A]\right|_{\{x(t)=x(0)\}\rightarrow\{x_\mu(t)=x_\mu(0)+n\beta w_\mu\}}
\end{equation} 
We use the Fourier transform of $A(x)=\sum \epsilon_\mu^i e^{ik_i x}$. The two 
point function can be obtained by deriving the effective action twice with 
respect to $\epsilon$. It 
is also necessary to integrate the so called zero mode 
\cite{Schub1} by making the change of variable: $x^\mu(\tau)= x^\mu_o+
n\beta\frac{\tau}{t}w^\mu+y^\mu(\tau)$ where $x_o$ is $\tau$
independent. By integrating the zero mode, the two point function can be written as 
\begin{eqnarray} 
\Pi^{\mu\nu}(k,\beta)&=&
-\frac{e^2}{2}\sum\limits_{n=-\infty}^{\infty}\int\limits^{\infty}_{0}\frac{dt}{t}[4\pi t]^{-2}\int\limits_0^t d\tau_1\int\limits_0^t
d\tau_2\nonumber\\ &&\exp\left(
-\frac{n^2\beta^2}{4t}+\frac{in\beta}{t} k_o (\tau_1-\tau_2)
\right)\exp(-k^2G_{12})\Big\{ -k^\mu k^\nu \dot{G}_{12}^2 \nonumber\\
&& \left. +i\frac{n\beta}{t}(w^{\mu }k^\nu +w^\nu
k^\mu)\dot{G}_{12}+g^{\mu\nu}\ddot{G}_{12}+w^\mu
w^\nu\frac{n^2\beta^2}{t^2}\right\} .
\label{eq:finitetemp}
\end{eqnarray}
 Where $G_{12}=G(\tau_1-\tau_2)$, is the translationally invariant propagator 
given in Eq.~(\ref{eq:green}).\\ 
The zero temperature result in Eq.~(\ref{eq:zerotemp}) can be obtained from 
Eq.~(\ref{eq:finitetemp}) by first putting $n=0$, then performing a partial 
integration to eliminate $\ddot{G}$. Hence from now on we subtract the $n=0$ 
contribution to obtain the finite temperature part of the two point 
function.\\ 
To simplify the $\tau_1$ and $\tau_2$ integration write
\begin{equation}
\Pi^{\mu\nu}(k,\beta)\equiv \sum\limits_{n=-\infty}^{\infty}\int\limits_0^t 
d\tau_1\int\limits_0^td\tau_2f_n(\tau_1-\tau_2),
\end{equation}
where $f_n(\tau_1-\tau_2)$ is the $n$-dependent expression in 
Eq.~(\ref{eq:finitetemp}). Using the symmetry properties of the propagator 
and its derivatives, it is easy to show that the sum over $n$ and the integral 
over $\tau_1$ and $\tau_2$ can be replaced by: 
\begin{equation}
\sum\limits_{n=-\infty}^{\infty}\int\limits_0^t d\tau_1\int\limits_0^t
d\tau_2
f_n(\tau_1-\tau_2)=2t^2\sum\limits_{n=-\infty}^{\infty}\int\limits_0^1du(1-u)f_n(ut),
\end{equation} 
where we have used the property $f_{-n}=f_n^*$.\\
Using the above transformation we obtain 
\begin{eqnarray}
\Pi^{\mu\nu}(k,\beta)&=&
-\frac{e^2}{(4\pi)^2}\sum\limits_{n=-\infty}^{\infty}\int\limits^{\infty}_{0}\frac{dt}{t}\int\limits_0^1
du(1-u)\nonumber\\ &&\hspace{-1.2cm}\exp\left(
-\frac{n^2\beta^2}{4t}+in\beta k_o u -k^2tu(1-u)\right)\Big\{ -k^\mu
k^\nu(1-2u)^2 \nonumber\\ &&
\hspace{-1.5cm}\left. +i\frac{n\beta}{t}(w^{\mu }k^\nu +w^\nu
k^\mu)(1-2u)+\frac{2}{t}g^{\mu\nu}(\delta(u)-1)+w^\mu
w^\nu\frac{n^2\beta^2}{t^2}\right\} .
\label{eq:mckeon} 
\end{eqnarray}
We note that Eq.~(\ref{eq:mckeon}) coincides with Eq.~(11) of ref. \cite{Mckeo1}.\\
It is important at this level to compare our derivation with the one used in 
\cite{Mckeo1}. The author of \cite{Mckeo1} has claimed that one could
not use the translationally invariant Green's function of 
Eq.~(\ref{eq:green}) at finite temperature. He claimed that such a Green's 
function is obtained by partial integration which will give boundary terms at 
finite temperature. Our derivation has 
shown that any expected boundary term, resulting from partial integration, 
cancels out. This allows the use of the translationally invariant Green's function. \\  
We proceed in a way similar to \cite{Mckeo1} to eliminate the $\delta$ 
function in  Eq.~(\ref{eq:mckeon}). Then by performing partial integration 
over $u$ we obtain
\begin{eqnarray} \Pi^{\mu\nu}(k,\beta)&=&
-\frac{e^2}{(4\pi)^2}\sum\limits_{n=-\infty}^{\infty}\int\limits^{\infty}_{0}\frac{dt}{t}\int\limits_0^1du\nonumber\\
&&\exp\left(-\frac{n^2\beta^2}{4t}+in\beta k_o u -k^2tu(1-u)
\right)\Big\{ (g^{\mu\nu}k^2-k^\mu k^\nu)(1-2u)^2\nonumber\\ &&
\left. +i\frac{n\beta}{t}(w^\mu k^\nu +w^\nu
k^\mu-g^{\mu\nu}w.k-\frac{k^2}{k_o}w^\mu w^\nu)(1-2u)\right\}.
\label{eq:2point} 
\end{eqnarray} 
It is necessary to verify gauge invariance of the above two point function. 
This is done in the next section. 
\subsection{Transversality} 
It is easy to verify transversality by contracting the two point 
function 
with $k_\mu$. To illustrate the transversality
property, let us define ${\cal P}_{\mu\nu}$ as:
\begin{equation} \frac{k^2}{k_o}{\cal P}^{\mu\nu}(k)=w^\mu k^\nu
+w^\nu k^\mu-g^{\mu\nu}w.k-\frac{k^2}{k_o}w^\mu w^\nu .
\end{equation} This tensor can be related to the photon transverse and
longitudinal projectors by (see appendix \ref{app:projector} for the 
definition of the projectors) 
\begin{equation} {\cal P}^{\mu\nu}= -{\cal
P}^{\mu\nu}_L -\frac{k_o^2}{k^2}{\cal P}^{\mu\nu}_T.  
\label{eq:proj}
\end{equation} 
The transverse and the longitudinal projectors are both
transverse to $k^\mu$. Besides the transverse projector is transverse
with respect to the 3-vector ${\bf k}$. Hence the two point function could be 
written in terms of ${\cal P}_{\mu\nu}$ and 
$g^{\mu\nu}-\frac{k^\mu k^\nu}{k^2}$ which are transverse. Therefore the two 
point function is manifestly transverse. 

\section{World-line formalism and Feynman parameters}
The world-line formalism and Feynman parametrization share the same 
property of reducing any n-point Green's function into a one loop integration. However 
it is well known that the Feynman parametrization is very subtle at finite 
temperature. In what follows we will show that the variable $u$ in 
Eq.~(\ref{eq:2point}) is analogous to a Feynman parameter. Then we figure out 
the above mentioned subtleties of Feynman parametrization and how they affect 
the calculation of the two point function.

\subsection{Rewriting the two point function}
We aim at rewriting the two point function in a form similar to a two point 
function evaluated using Feynman parameters.
Doing partial integration on t in Eq.~(\ref{eq:2point}) we get rid of the $1/t^2$ and we obtain 
\begin{eqnarray}
\Pi^{\mu\nu}(k,\beta)&&= -\frac{e^2}{(4\pi)^2}\sum\limits_{n=-\infty}^{\infty}\int\limits^{\infty}_{0}\frac{dt}{t}\int\limits_0^1du(1-2u)\nonumber\\
&&\exp\left(-\frac{n^2\beta^2}{4t}+in\beta k_o u -k^2tu(1-u) \right)\Big\{- \frac{k^2{\bf k}^2}{k_o^2}{\cal P}^{\mu\nu}_L(1-2u)\nonumber\\
&&\left .+\frac{2}{in\beta k_o}\left(\frac{k^2}{k_o^2}\right)^2{\cal P}^{\mu\nu}\right\}+ \frac{3m^2}{2}\frac{k^2}{k_o^2}{\cal P}^{\mu\nu}.
\end{eqnarray} 
The last term is a boundary term resulting from partial integration. The 
thermal mass is defined as $m^2=e^2T^2/3$.
The t-integration gives a Bessel function of second type \cite{GradsR1}
\begin{equation}
\frac{1}{2}\int\limits_0^\infty \frac{dt}{t}\exp\left[- \frac{x}{2}(t+\frac{z^2}{t})\right]\equiv K_{0}(xz) ,
\end{equation}
where $x=2k^2u(1-u)$ and $z=n\beta/2\sqrt{u(1-u)k^2}$. By using the 
transformation $n\rightarrow -n$ for negative $n$ in the sum we obtain
\begin{eqnarray}
\Pi^{\mu\nu}(k,\beta)&&= -\frac{4e^2}{(4\pi)^2}\sum\limits_{n=1}^{\infty}\int\limits_0^1du(1-2u)K_{0}(n\beta\sqrt{u(1-u)k^2})\nonumber\\
&&\times\Big\{- \cos(n\beta k_ou)\frac{k^2{\bf k}^2}{k_o^2}{\cal P}^{\mu\nu}_L(1-2u) \nonumber\\
&& +2\left(\frac{k^2}{k_o^2}\right)^2{\cal P}^{\mu\nu}\int\limits_0^u dx\cos(n\beta k_o u)\Big\}+ \frac{3m^2}{2}\frac{k^2}{k_o^2}{\cal P}^{\mu\nu}.
\label{eq:2point-t}
\end{eqnarray} 
The next step is to evaluate the sum over $n$. This is done via the method 
proposed in appendix \ref{app:sum}. Using Eqs.~(\ref{eq:sum-finite})-(\ref{eq:square}), 
and performing the $x$-integration in the above expression, the finite 
temperature dependence of the two point function is found to be    
\begin{eqnarray}
\Pi^{\mu\nu}_{\beta}(k)&=&\frac{3m^2}{2}\frac{k^2}{k_o^2}{\cal P}^{\mu\nu}-\frac{e^2}{2\pi^2}T\sum\limits_{m=-\infty}^{\infty}\left\{-\pi\frac{k^2}{k_o^2}{\cal P}^{\mu\nu}\sqrt{(\omega_m-k_o)^2}\right.\nonumber\\
&&\!\!\!\!\!\!\!\!\!\!\!\!+\int\limits_0^\infty dy y^2\int\limits_0^1\frac{du }{[y^2-{\bf k}^2u^2+[k^2-2k_o\omega_m]u+\omega_m^2]^2}\nonumber\\
&&\!\!\!\!\!\!\!\!\!\!\!\!\!\!\times \Big[-\frac{k^2{\bf k}^2}{k_o^2}{\cal P}^{\mu\nu}_L(1-2u)^2\left. +2u\frac{k^2}{k_o^2}{\cal P}^{\mu\nu}(k^2-2k_o\omega_m-2{\bf k}^2u)\Big]\right\}.
\label{eq:2point-}
\end{eqnarray}
Where $\omega_m=2m\pi T$ are the standard Matsubara frequencies for bosons. 
A further simplification could be done by integrating  
by part the term proportional to ${\cal P}^{\mu\nu}$ in the above equation  
\begin{eqnarray}
\Pi^{\mu\nu}_{\beta}(k)&=&-\frac{e^2}{2\pi^2}T\sum\limits_{m=-\infty}^{\infty}\int\limits_0^\infty dy y^2\int\limits_0^1\frac{du }{[y^2-{\bf k}^2u^2+[k^2-2k_o\omega_m]u+\omega_m^2]^2}\nonumber\\
&&\times \Big[-\frac{k^2{\bf k}^2}{k_o^2}{\cal P}^{\mu\nu}_L(1-2u)^2 +\frac{4y^2}{3}\frac{k^2}{k_o^2}{\cal P}^{\mu\nu}\Big]+ \frac{3m^2}{2}\frac{k^2}{k_o^2}{\cal P}^{\mu\nu}.
\end{eqnarray}
This expression is analogous to a two point function, evaluated at one loop 
level, in the imaginary time formalism after the application of 
Feynman parameterization (the $u$ integral). To determine the momenta of the 
particles in the loop we can consider $y$ as the magnitude 
of a vector ${\bf y}$. Then we perform the following change of variables 
${\bf p}={\bf y}+u{\bf k}$. Having a $y^2$ dependent expression for the two 
point function, it is then possible to add any odd function of ${\bf y}$. 
Using Eq.~(\ref{eq:proj}) and adding the suitable odd function of ${\bf y}$ 
to eliminate the $u$ dependence in the numerator we obtain     
\begin{eqnarray}
\Pi^{\mu\nu}_{\beta}(k)&=& \frac{3m^2}{2}\frac{k^2}{k_o^2}{\cal P}^{\mu\nu}+2e^2T\sum\limits_{m=-\infty}^{\infty}\int \frac{d^3{\bf p}}{(2\pi)^3}\int\limits_0^1\frac{du}{[P^2(1-u)+(k-P)^2u]^2}\nonumber\\
&&\times \Big\{({\bf p}.{\bf p}-\frac{({\bf p}.{\bf k})^2}{{\bf k}^2}){\cal P}^{\mu\nu}_T+\frac{k^2}{2{\bf k}^2k_o^2}\left[({\bf p}-{\bf k})^2-{\bf p}^2\right]^2{\cal P}^{\mu\nu}_L\Big\}
\label{eq:feynman}
\end{eqnarray}      
with $P\equiv(\omega_m,{\bf p})$.\\ 
The strict application of world-line formalism indicates that we could 
perform the $u$-integration. Then we can do analytic continuation to obtain 
retarded or advanced two point function. This approach gives wrong results. To 
see this, note that the parameter $u$ plays the role of a Feynman parameter 
applied for the product of two propagators of momentum $P$ and $(P-k)$. As 
we will show in the next section the naive application of Feynman parametrization 
leads to wrong results at finite temperature. To avoid the  $u$-integration we `undo'  Feynman 
parametrization. It is then straightforward to show that the two point 
function could be written as
\begin{equation}
\Pi^{\mu\nu}_{\beta}(k)= -\frac{3m^2}{2}g^{\mu\nu} + e^2T\sum\limits_{m=-\infty}^{\infty}\int \frac{d^3{\bf p}}{(2\pi)^3}\frac{4P^\mu P^\nu -k^\mu k^\nu}{P^2(P-k)^2}
\label{eq:twopoint}
\end{equation}
which coincides with the expression obtained directly in the imaginary time 
formalism. The first term on the right hand side of Eq.~(\ref{eq:twopoint}) is the contribution of the 4-photon vertex, and the 
second term comes from the photon-scalar coupling. In the imaginary 
time formalism there is a systematic way of doing analytic continuation to 
obtain: retarded, advanced or Feynman two point function
\begin{eqnarray}
\Pi^{\mu\nu}_{R}(k_o)&=&\Pi^{\mu\nu}_{\beta}(k_o+i\epsilon)\nonumber\\
\Pi^{\mu\nu}_{A}(k_o)&=&\Pi^{\mu\nu}_{\beta}(k_o-i\epsilon)\nonumber\\
\Pi^{\mu\nu}_{F}(k_o)&=&\Pi^{\mu\nu}_{\beta}(k_o+i\epsilon k_o).
\end{eqnarray}    
However this method renders the world-line formalism useless. Since after a 
huge computational effort we get an expression, for the two point function, 
which is obtained by direct application of Feynman's rules in the imaginary 
time formalism.\\
Therefore: is it possible to cure the analytic structure of 
Eq.~(\ref{eq:feynman}) with the possibility of doing directly the 
$u$-integration? In the next section we try to give an answer to this question.

\subsection{Feynman parameters at finite temperature}
As we have shown in the previous section the world-line formalism incorporate 
the use of Feynman parameter method. The Feynman parameter method has a potential problem at 
finite temperature, and misleading results could be obtained \cite{BedaqD1}. The basic difference 
between the zero and finite temperature field theory is that, in the former it 
is possible to write any Green's function entirely in terms of retarded or 
advanced propagators. At finite temperature one can not avoid the appearance 
of the product of retarded and advanced propagators. Hence we could have the 
following combination
\begin{equation}
\frac{1}{a+i\epsilon}\frac{1}{b-i\epsilon}
\end{equation}
where $\epsilon$ is an infinitesimal positive parameter. Applying ``naive'' 
Feynman parameterization to this product will give:
\begin{equation}
\frac{1}{a+i\epsilon}\frac{1}{b-i\epsilon}={\cal P}\int\limits_0^1\frac{dx}{[a(1-x)+bx+i\epsilon(1-2x)]^2}
\end{equation}
the imaginary part of the denominator vanishes at $x=1/2$. In ref. 
\cite{Weldo1} Weldon has shown that the correct way to do a Feynman integral is:
\begin{equation}
\frac{1}{a+i\epsilon}\frac{1}{b-i\epsilon}={\cal P}\int\limits_0^1\frac{dx}{[a(1-x)+bx+i\epsilon(1-2x)]^2}+\frac{4\pi i \delta(a+b)}{a-b+2i\epsilon}.
\label{eq:weldon}
\end{equation}  
This modification renders the use of Feynman parameterization very tricky. To 
obtain the correct analytic structure one should add to the two point function 
a very peculiar function, resulting from the $\delta$ term in 
Eq.~(\ref{eq:weldon}). \\
We come now to the question asked in the previous section. Having an 
'underlying' use of Feynman parametrization, the world line formalism suffers 
from the same subtleties mentioned above. Hence a naive use of the world-line 
formalism can not give the correct retarded or advanced two point function. 
To obtain the correct analytic structure of the two point function one 
should correct the naive $u$-integration in Eq.~(\ref{eq:feynman}) by adding 
the same function proposed by Weldon in \cite{Weldo1}. However this is far 
from being systematic, and can not be generalized to any n-point Green's 
function.

\section{Summary and outlook}
In this paper we have studied the analytic features of the world-line 
formalism. We have shown that a translationally invariant 
Green's function could be used at finite temperature. However this 
formalism has an intrinsic problem of analytic continuation. To solve this 
problem we can proceed in one of the following ways: 
\begin{itemize}
\item Rewrite the two point function, derived in the world line formalism, to 
obtain an expression similar to the one obtained in the imaginary time 
formalism. Hence it does not worth using the world-line formalism, and a 
simpler method will be the direct application of the imaginary time formalism. 
\item We could correct the analytic structure of the two point function by 
adding an {\it adhoc} function. However this method is not systematic. We can 
not implement such solution to obtain a self consistent world-line formalism 
at finite temperature.  
\end{itemize}
Our study has shown that the naive application of the world-line formalism at 
finite temperature gives wrong results for the two point function. Hence a 
real time approach to the world-line 
formalism would have save us from the acrobatic way of obtaining the correct 
analytic behavior of any Green's function. However in the presence of 
external field the correct analytic structure could be obtained by 
calculating directly the two point function with this field. 
   
\section{Acknowledgment}

I would like to thank C. Schubert for fruitful discussions. I would like also 
to thank E. Pilon for careful 
reading of this manuscript. I also thank the LAPTH for their hospitality.

\appendix

\section{projectors}
\label{app:projector}
In the rest frame of the plasma, the transverse and longitudinal projectors 
are given by \cite{LandsW1}
\begin{eqnarray}
P^{^{T}}_{\mu o}(k)=0 \qquad ,\;\;  P^{^{T}}_{ij}(k)=\delta_{ij} -
{\frac{k_i k_j}{{\bf k}^2}}
\nonumber \\
P^{^{L}}_{\rho\sigma}(k)=
  -P^{^{T}}_{\rho\sigma}(k)+g_{\rho\sigma}-{{k_\rho k_\sigma}\over{k^2}}
  \; . 
\end{eqnarray}
Those projectors are transverse with respect to the photon momentum $k$
$$ k_\mu P^{\mu\nu}_{_{T,L}}(k)=0.$$
 
\section{Summation over $n$}
\label{app:sum}
The sum over the winding number $n$ could be done using the following sum rule 
of the Bessel function $K_0$ \cite{GradsR1}
\begin{eqnarray}
\sum\limits_{n=1}^{\infty}\cos(n\alpha)K_0(n\phi)&=&\frac{1}{2}\left(\gamma_{\rm E}+\ln{\frac{\phi}{4\pi}}\right) +\frac{\pi}{2\sqrt{\phi^2+\alpha^2}}\nonumber\\
&&+\frac{1}{2}\sum\limits_{n=1}^\infty{\left[\frac{1}{\sqrt{\phi^2+(2m\pi-\alpha)^2}}-\frac{1}{2m\pi}\right]}\nonumber\\
&&+\frac{1}{2}\sum\limits_{n=1}^\infty{\left[\frac{1}{\sqrt{\phi^2+(2m\pi+\alpha)^2}}-\frac{1}{2m\pi}\right]}
\end{eqnarray} 
Where $\gamma_{\rm E}$ is the Euler constant.
By substituting $\alpha=\beta k_o u$ and $\phi=\beta\sqrt{u(1-u)k^2}$ the sum 
over $n$ in the two point function could be done. Being interested in the 
finite temperature part we can keep the $\beta$ dependent terms of the above 
sum:
\begin{eqnarray}
\sum\limits_{n=1}^{\infty}\cos(n\alpha)K_0(n\phi)|_\beta &&=\frac{\pi}{2\sqrt{\phi^2+\alpha^2}}+ \frac{1}{2}\sum\limits_{n=1}^\infty\frac{1}{\sqrt{\phi^2+(2m\pi-\alpha)^2}}\nonumber\\
&&+\frac{1}{2}\sum\limits_{n=1}^\infty\frac{1}{\sqrt{\phi^2+(2m\pi+\alpha)^2}}\nonumber\\
&&\hspace{-6mm}=\frac{\pi T}{2}\sum\limits_{m=-\infty}^{\infty}\frac{1}{\sqrt{-{\bf k}^2u^2+[k^2-2k_o\omega_m]u+\omega_m^2}}
\label{eq:sum-finite}
\end{eqnarray} 
where $\omega_m=2\pi mT$ are the Matsubara frequencies. A further simplification 
can be done if we transform the square root into a simple fraction. This can 
be achieved using 
\begin{equation}
\int\limits_0^\infty \frac{y^2dy}{(y^2+a)^2}=\frac{\pi}{4\sqrt{a}} .
\label{eq:square}
\end{equation}

\end{document}